\documentclass{elsart}

\usepackage{natbib}

\usepackage{graphicx}
\usepackage{amssymb}

\begin{document}

\begin{frontmatter}


\title{The ASTRO-F Mission : Large Area Infrared Survey}

\author[label1]{Hideo Matsuhara\corauthref{cor1}}
\ead{maruma@ir.isas.jaxa.jp}
\corauth[cor1]{}
\author[label2]{Hiroshi Shibai}
\author[label3]{Takashi Onaka}
\author[label1]{Fumihiko Usui}
\address[label1]{Department of Infrared Astrophysics, Institute of Space
and Astronautical Science (ISAS), Japan Aerospace Exploration Agency (JAXA), 
3-1-1 Yoshinodai, Sagamihara, Kanagawa 229-8510, Japan}
\address[label2]{Graduate School of Science, Nagoya University, Furo-cho,
Chikusa-ku, Nagoya 464-8602, Japan}
\address[label3]{Department of Astronomy, University of Tokyo, 7-3-1 Hongo,
Bukyo-ku, Tokyo 113-0033, Japan}

\begin{abstract}
ASTRO-F is the first Japanese satellite mission dedicated for large
area surveys in the infrared. The 69cm aperture telescope and scientific
instruments are cooled to 6K by liquid Helium and mechanical coolers.
During the expected mission life of more than 500 days, ASTRO-F will
make the most advanced all-sky survey in the mid- to far-infrared since 
the Infrared astronomical Satellite (IRAS).
The survey will be made in 6 wavebands and will include the first all sky survey at 
$>$100-160~$\mu$m.
Deep imaging and spectroscopic surveys with pointed observations will also
be carried out in 13 wavelength bands from 2-160~$\mu$m. ASTRO-F
should detect more than a half million galaxies tracing the large-scale
structure of the Universe out to redshifts of unity, detecting rare,
exotic extraordinarily luminous objects at high redshift, numerous brown
dwarfs, Vega-like stars, protostars, and will reveal the large-scale
structure of nearby galactic star forming regions. ASTRO-F
is a perfect complement to Spitzer Space Telescope in respect of
its wide sky and wavelength coverage. Approximately 30 percent of pointed
observations will be allocated to an open-time opportunity. Updated pre-flight
sensitivities as well as the observation plan including the large-area surveys
are described.
\end{abstract}


\end{frontmatter}

\section{Introduction}

The Japanese infrared astronomy mission ASTRO-F is a second
generation infrared all-sky survey mission following the Infrared Astronomical 
Satellite (IRAS, Neugebauer et al. 1984) with much improved sensitivities, 
spatial resolutions and wider wavelength coverage.
ASTRO-F is equipped with a 69cm cooled (approximately to 6K) Ritchey-Chretien 
type telescope and two focal plane instruments covering the 2-160$\mu$m wavelength
range. In August 2003, NASA launched the Spitzer 
Space Telescope (SST), an 85cm cooled telescope with three focal plane 
instruments covering the 3-160$\mu$m wavelength range (Werner et al. 2004).
ASTRO-F will be complementary to SST with respect to having wider sky coverage
and continuous wavelength coverage. ASTRO-F is now scheduled
to be launched in January-February 2006 by the M-V rocket of JAXA 
(Japan Aerospace Exploration Agency), and it is expected that
the mission will provide an invaluable database for
planning of observations with the Herschel Space Observatory (Pilbratt 2003).

The purpose of this paper is to present the updated pre-flight sensitivities
in various observing modes, and to describe the current observation plan with 
emphasis on the large-area surveys. Details of the ASTRO-F mission and 
satellite system are described in Murakami (2004).
Further details on the current performance of the focal plane instruments are
described in Kawada et al. (2004) and Onaka et al. (2004).

\section{Overview of ASTRO-F Mission}

   \subsection{Orbit and attitude control}

\begin{figure}[htbp]
\begin{center}
\includegraphics[width=14cm,clip]{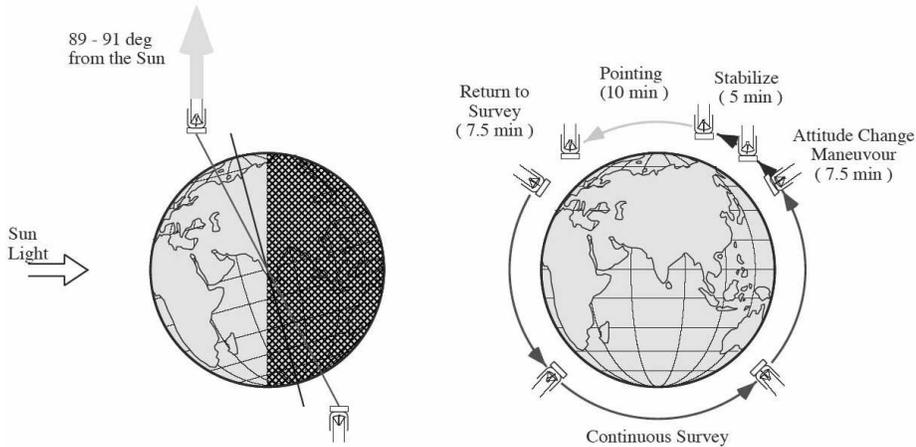}
\caption{The ASTRO-F spacecraft attitude is schematically shown. (Left) in
the all-sky survey mode,
the spacecraft rotates uniformly around the axis directed toward the Sun once
every orbital revolution, resulting in a continuous  scan of the sky. The
whole sky can be covered in a half year. (Right) the attitude operation for
the pointing observations.  Due to the
limit imposed by the earthshine illumination to the telescope baffle, the 
duration
of the pointing is
limited to 10 minutes. The pointing direction can be freely chosen in the
telescope orbital plane given by the survey mode attitude, however is
restricted to within $\pm 1$deg in the direction perpendicular to the orbital
plane.}
\label{orbit}
\end{center}
\end{figure}

\begin{figure}[htbp]
\begin{center}
\begin{minipage}{6.9cm}
%\hspace{-3mm}
\includegraphics[width=6.7cm]{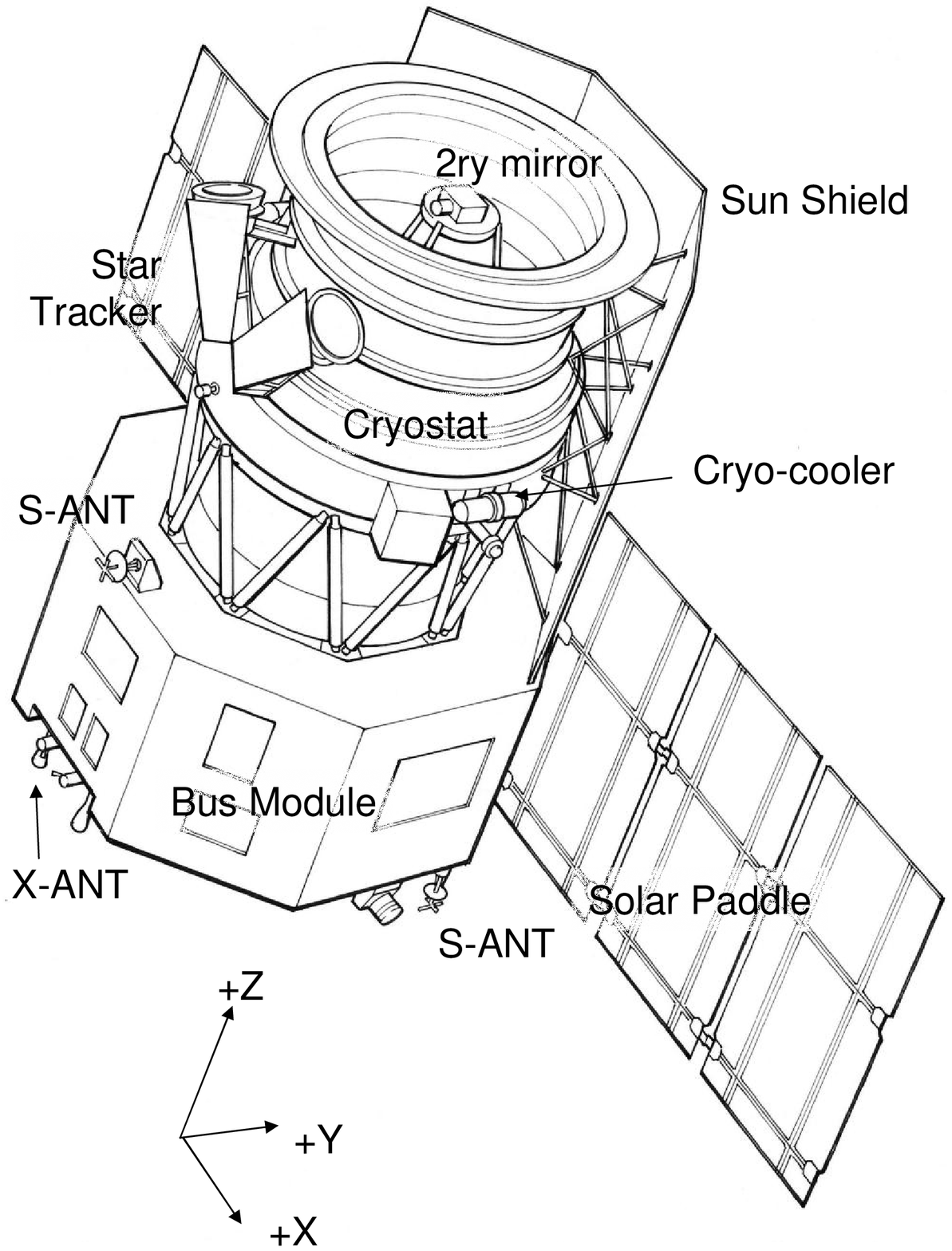}
\caption{The ASTRO-F spacecraft (2m $\times$ 1.9m $\times$ 3.7m at launch). The
weight is 960kg including fuels for the reaction control system.}
\label{spacecraft}
\end{minipage}
\begin{minipage}{6.9cm}
%\hspace{3mm}
\includegraphics[width=6.7cm,clip]{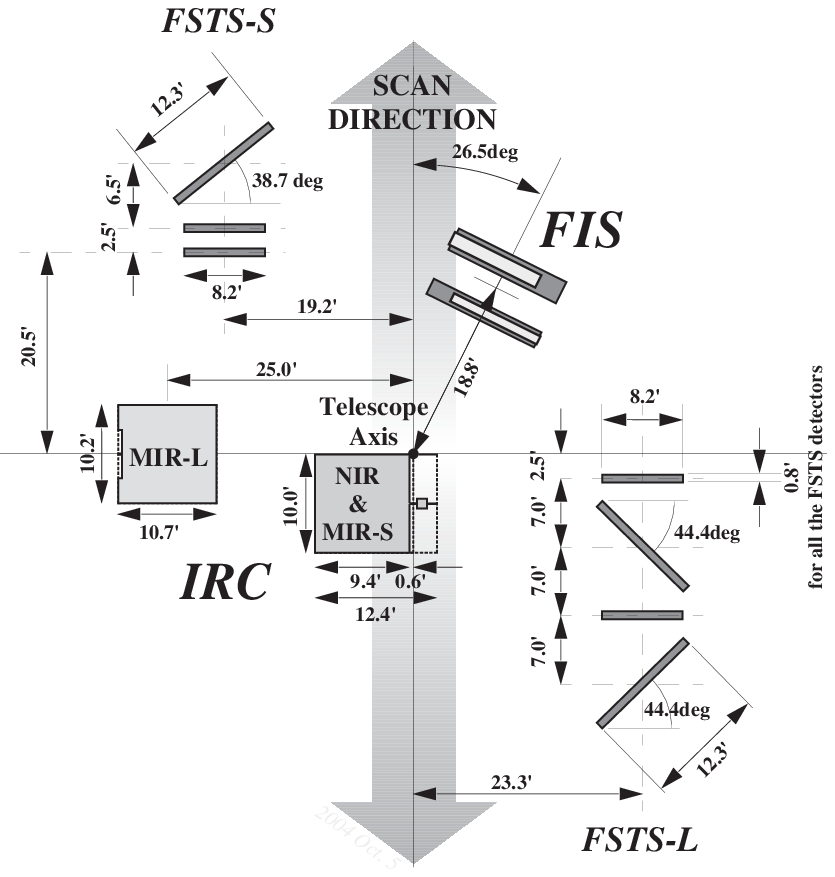}
\caption{Field-of-view configuration of ASTRO-F instruments, projected on
the sky.}
\label{fov}
\end{minipage}
\end{center}
\end{figure}

The orbit of ASTRO-F is a sun-synchronous polar orbit at an altitude of 750km and
at an inclination of 98.4 deg. The spacecraft +Y-axis (shown in Figure~\ref{spacecraft})
is always pointed in the direction of the Sun, thus avoiding the sunlight
illuminating the telescope baffles.
While orbiting around the Earth with a period of approximately 100 minutes,
the ASTRO-F spacecraft spins around the Y-axis avoiding the earthshine from
illuminating the telescope. Thus, this attitude ensures a stable thermal condition of the
spacecraft, and moreover results in a continuous scan of the sky 
with a speed of 3.6 arcminutes per minute. Figure~\ref{orbit}
(left) schematically shows the attitude of ASTRO-F in this all-sky 
survey mode.
A second method of operation is the pointing mode, where the telescope observes a 
certain sky position for a longer exposure with the focal-plane instruments 
(Figure~\ref{orbit} right). Due to the limitations required to prevent earthshine
from illuminating the telescope baffle, the duration of the pointing mode of operation is 
limited to 10 minutes.
The visibility zone({\it i.e.} allowed peak-to-peak angle of the pointing 
direction in the cross-scan
direction) is $\pm 1$deg, which is determined by the field-of-view of the 
onboard fine sun sensor and by the design of the baffle and the sun shield. 
The attitude control system provides additional
capabilities to slightly change the pointing direction for the pointing modes:
step-scan, slow-scan, and micro-scan. The slow-scan is a uniform scan with a much 
slower scan speed(less than 30 arcseconds per second) than that of the all-sky
survey. The slow-scan survey is especially useful for covering a much larger sky 
areas than those obtainable by pointing, to much better sensitivities than that 
of the all-sky survey.
However, the overheads for a pointing attitude mode are significantly large: 
approximately 20 minutes is required for the maneuvering and stabilization. Hence 
the number of pointing observations in one orbit is restricted to three or less.

   \subsection{Scientific instruments}

ASTRO-F incorporates two focal plane instruments covering the infrared wavelength from
2 to 160~$\mu$m. The first focal plane instrument is the Far-Infrared Surveyor 
(FIS, Kawada et al. 2004) which will primarily survey the entire sky simultaneously in 
four far-infrared bands from 50 to 160~$\mu$m with approximately
diffraction-limited spatial resolutions (30-60 arcsec).
The second focal-plane instrument is the InfraRed Camera (IRC, Wada et al. 2003; 
Onaka et al. 2004), which is designed to take deep images of selected sky 
regions by pointed observations from 2 to 26~$\mu$m.
Both instruments are equipped with spectroscopic capabilities with 
moderate or low resolution, however, they are beyond the scope of this paper.
Figure~\ref{fov} shows the field-of-view configuration on the sky, and the waveband
characteristics for broad-band imaging are summarized in Table~\ref{tab_imaging}.
The pre-flight sensitivities estimated from the current laboratory 
experiments of the telescope and the instruments at low temperature are shown in 
Table~\ref{tab_survey}, \ref{tab_pointing}, and \ref{tab_slow}.

\section{Large Area Surveys with ASTRO-F}

The primary objective of the ASTRO-F mission is to make an all-sky survey 
in 6 wavebands
covering 9-160$\mu$m. In addition, large-area {\it Legacy} surveys in pointing mode
are also planned in two dedicated fields near the ecliptic poles: the north
ecliptic pole (NEP) itself and the Large Magellanic Could (LMC) near the south 
ecliptic pole (SEP).

   \subsection{All-sky survey}

\begin{figure}[htbp]
\begin{center}
\includegraphics[width=10cm,clip]{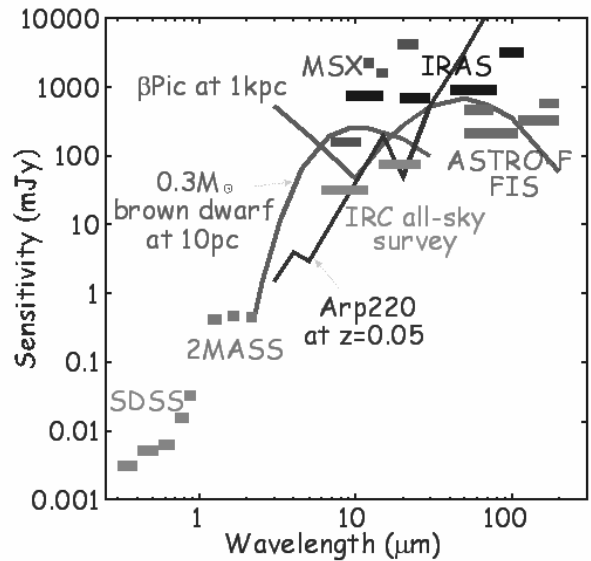}
\caption{Flux limit sensitivities (for one scan, 5$\sigma$) of the ASTRO-F
all-sky survey compared with those of large-area surveys at other wavelengths. }
\label{fl_survey}
\end{center}
\end{figure}

\begin{figure}[htbp]
\begin{center}
\includegraphics[width=14cm,clip]{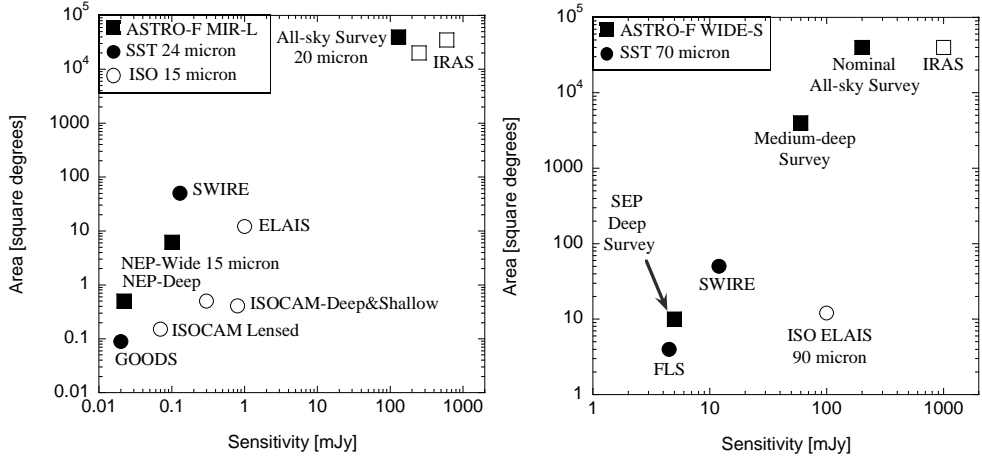}
\caption{Survey depth vs areal coverage comparison of ASTRO-F large-area
survey with surveys with
    past/on-going infrared astronomical satellites: (right) far infrared,
(left) mid infrared. }
\label{deptharea}
\end{center}
\end{figure}

In the all sky survey mode, the telescope beam continuously scans along a
great circle perpendicular to the direction of the Sun with a period of 100 
minutes(see Figure~\ref{orbit}).
Hence the survey paths are nearly aligned to the ecliptic lines of longitude,
with approximately 4.1 arcmin spacing at the ecliptic plane
between successive orbits. Since the cross-scan width of the short-wave channel of 
the FIS is 8.2arcmin, the single ``hours confirmation" can be established at every 
point of the sky, as long as successive orbits are useful for observations without, 
for example, the presence of the South Atlantic Anormaly or the Moon. 
In phase-1, most of the sky (approximately 85\%-95\%, depending on the
observation scenario, see next section) will be observed in 6 wavebands covering
9--160$\mu$m (Table~\ref{tab_imaging} and \ref{tab_survey}).
supplementary survey observations
will be made in phase-2 in order to fill the gaps in the phase-1 sky coverage.
The flux limit (for one scan) of the ASTRO-F all-sky survey is
shown in Figure~\ref{fl_survey}
compared with those of large-area surveys at other wavelengths. The
uniqueness of the ASTRO-F all-sky
survey lies in its' wider wavelength coverage(up to 160$\mu$m) and
much finer spatial resolution compared to that of IRAS. Moreover, thanks to the
high visibility in the regions near the ecliptic poles, we will also be 
able to obtain a faint source
catalog with fluxes down to 80mJy at 90$\mu$m (5$\sigma$) over a substantially
large area($\sim$4000 deg$^{2}$), which will complement  the on-going 
SWIRE(Lonsdale et al. 2004)
survey with SST (see Figure~\ref{deptharea}(right)).
ASTRO-F in all-sky survey mode will detect approximately half million
galaxies tracing the
large-scale structure of the Universe out to redshifts of unity, detecting rare,
exotic extraordinarily luminous objects at high redshift, numerous brown
dwarfs, Vega-like stars, protostars, and will reveal the large-scale
structure of the star-forming regions in both the solar neighbourhood, 
the distribution of the interplanetary dust in our solar system, and
towards the Galactic plane
where SST cannot observe because of the saturation of its' detectors (30-50
MJy/sr at 160$\mu$m).

   \subsection{NEP survey}

\begin{figure}[htbp]
\begin{center}
\includegraphics[width=14cm,clip]{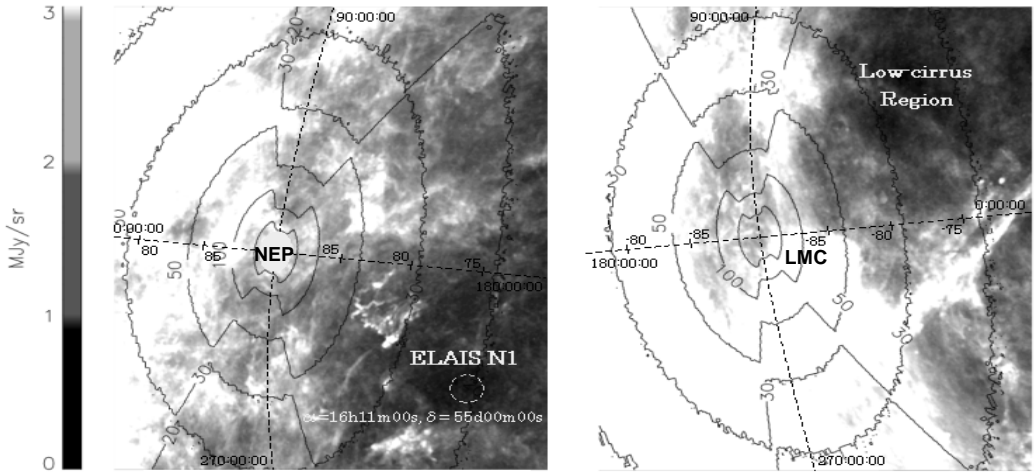}
\caption{ASTRO-F target fields of the large-area survey programs in
pointing mode super-imposed on the
IRAS 100$\mu$m images. The ASTRO-F visibility contours for phases 1
and 2 are indicated
by solid lines. (Left) the field in the NEP.
6.2~deg$^2$ survey for
mainly extragalactic science is currently planned. Also ``ELAIS N1", a
SST/SWIRE field,
is indicated, since it is considered as a candidate for a
follow-up ASTRO-F survey at 11 and 15$\mu$m.
(Right) fields near the SEP. The LMC is the
main target for the stellar and interstellar matter studies, while the low-cirrus
region is also considered for an additional deep extragalactic survey.}
\label{nepsep}
\end{center}
\end{figure}

Due to the constraint of the orbit and the rather limited visibility zone 
(see the previous section), the sky visibility for the ASTRO-F telescope is 
strongly weighted toward the ecliptic poles. Hence a large area survey of more 
than $\sim$1 deg$^{2}$ in the pointing mode is only possible in regions close 
to the ecliptic poles. Figure~\ref{nepsep} shows the location of the survey 
fields currently planned in the pointing mode.

The main driver for the NEP survey is an unbiased survey of extragalactic
sources. the survey is divided into two realizations: ``NEP-Deep" and 
``NEP-Wide". The former is a deep survey consisting of 504 pointing observations
covering 0.5~deg$^{2}$. This is designed to obtain two major science goals:
the dusty star-formation history of the Universe out to $z=3$, and the mass
assembly history and evolution of large scale structure out to $z=4$. 
The IRC in the broad band imaging mode will be used for this survey,
with 9 wavebands covering 2.4 - 24$\mu$m. Deep optical and near infrared (up
to the $K_{\rm s}$-band) imaging data are already available in this field. 
The latter ``NEP-Wide" survey  is a 6.2~deg$^{2}$
shallow survey over a circular area centered near the NEP with 450 pointing 
observations in total. This survey is especially designed to complement the
 survey volume of the ``NEP-Deep" observing program for the $z \sim 1$ 
Universe ({\it i.e.} to 
limit uncertainties due to the significant cosmic variance in the $z \leq 1$ 
Universe). Nine IRC wavebands will be used for this survey. Inclusion of the N2
band (2.4$\mu$m) is mandatory for the study of spatial fluctuations of the cosmic near 
infrared background. The most important role of these surveys is to perform 
large-area 11 and 15$\mu$m surveys, which cannot be done with SST. The 15$\mu$m 
survey volume with ASTRO-F/IRC will far exceed that of ISO, generating a 
sample of more than ten thousand ULIRG/starburst galaxies. Thanks to the
continuous wavelength coverage of the ASTRO-F/IRC over the 8-24$\mu$m range, 
we will be able to estimate photometric redshifts by using the silicate/PAH
features, allowing us to construct accurate luminosity functions to $z=1-1.5$.
The depth in the mid-infrared and the sky coverage of the NEP survey are 
compared with other missions(ISO, SST) in Figure~\ref{deptharea}(left).

Figure~\ref{nepsep}(left) also shows the location of the ELAIS-N1 field, an 
on-going SST/SWIRE legacy survey field located at the highest ecliptic latitude,
 which is a candidate for a follow-up study with ASTRO-F.
  Except for the 10-20$\mu$m region, deep multi-wavelength data has already 
been accumulated and a follow-up study using solely the 11 and 15$\mu$m bands 
in the IRC will provide us with a complementary data set
to that expected from the ``NEP-Wide" survey.

As for the far infrared data at the NEP, since so many scan paths overlap at
the pole, the all-sky survey will reach the galaxy confusion limit. This 
depth is enough to yield the far-infrared detection
of $z \geq 2$ hyper-luminous infrared galaxies(with $\sim 10^{13}L_{\odot}$
infrared luminosity) which will be identified in the mid infrared with 
the ASTRO-F/IRC. However, such a faint source extraction
will be a very hard task and likely to be time consuming. We may therefore 
consider an additional slow-scan survey with the
FIS which only requires a few $\sim$10 pointing observations, to generate
deep far infrared images of the NEP rather quickly.

   \subsection{LMC survey}

The LMC is seen in nearly face-on, and its  proximity ($\sim$50kpc)
provides us with a unique opportunity
to resolve and separate individual components even in the far
infrared ($30"$ corresponds to 8pc at the distance of the LMC). The LMC
has been observed at many other wavelengths (X, UV, optical, near infrared,
$H_{\alpha}$, HI, CO). The LMC was also extensively observed in the mid infrared 
by the Midcourse Space Experiment (MSX) over 100~deg$^2$ area with a flux 
limit of 50~mJy at 8$\mu$m (Egan et al. 2001). Therefore, the LMC is one of the 
best target fields for the study of interstellar matter, star-formation, and 
late-type stars.
As an ASTRO-F large-area survey, an approximately 20 deg$^{2}$ fan-shape 
area will be surveyed in 6 IRC imaging bands covering 3.3 - 
26$\mu$m\footnote{The $7\times7$deg$^2$ LMC survey with SST has been approved 
in the GO-2 observations. Thus the ASTRO-F survey strategy may be slightly 
revised. }.
The survey area covers the molecular ridge, supernova remnants, and other
interesting regions.
These 6 band data can distinguish and identify the nature of detected
objects, such as proto-stellar objects, young stellar objects, and AGB stars.
The required number of pointing observations is 950 in total.
% *** the following 2 sentences are NOT relevant since SST will perform this!! ***
% SST could make a similar large-area survey toward the LMC, but two months 
% observing time has to be allocated.
The ASTRO-F/IRC survey will provide mid-infrared images with higher angular resolution
and sensitivities than those of MSX, and is complementary to  SST in 
its continuous wavelength coverage including the 11 and 15$\mu$m bands.
These bands are particularily useful in investigating
the formation and destruction processes of very small grains and PAHs.
The ASTRO-F data, when combined with SST ones, will address a
number of astrophysical problems, ranging the entire birth to death cycles of 
stars, and give us a first insight into the
detailed material circulation in a galaxy.

   \subsection{Other surveys near the SEP}

Since the LMC is not always observable during the mission, additional 
pointing surveys can be allocated somewhere near the
SEP, as long as this does not reduce the sky coverage of the all-sky survey.
The Chamaeleon molecular cloud
is relatively close to the pole($\beta$=60-65deg) and hence it is possible
to search for young stellar objects and proto-planetary disks extensively by
using a slow-scan survey
in both the IRC and FIS wavebands. In addition, as shown in Figure~\ref{nepsep}
(right), near the south ecliptic pole there is a low-cirrus
region (Shlegel et al. 1998) which will be an
optimum field for the study of the large-scale fluctuations of the cosmic
far infrared background. ASTRO-F/FIS is more suitable for
the study of such diffuse sources than SST/MIPS, thanks to its larger
pixel field-of-view. Moreover
cross-correlation analysis between the 4 FIS wavebands is a unique capability
of ASTRO-F. 

\subsection{Data products from the large-area survey}

Each large-area survey requires a dedicated team for the observation planning
and the data analysis. In particular for the all-sky survey, pipeline
development to generate the
point source catalogue has been on-going by an international data handling
team (Japan, Korea, United Kingdom,
the Netherlands). From the all-sky survey data, the following source
catalogues will be released:

\begin{itemize}
\item{ASTRO-F fluxes of known sources (especially IRAS PSC) : released in
the survey period.}
\item{Bright Source Catalogue(BSC): this is the primary all-sky survey
product with uniform flux limit (Table~\ref{tab_survey}) over the entire sky.
The catalogue release is planned to commence one month from the end of the 
survey and finished within one year after the end of the survey. }
\item{Faint Source Catalogue(FSC): supplemental catalogue of the fainter
sources in the regions with higher
redundancy ({\it i.e.} regions at high ecliptic latitudes). The FSC release
date will be at least one year after the release of the BSC. }
\end{itemize}

The catalogues will be initially released to the ASTRO-F team, and then 
open to the public one year later.
Image data, especially on small scales, will also be released, although
more investigation is  still
necessary to define the data release plan.

Regarding the data products of the NEP and the LMC surveys, the BSC as well
as FSC are also
generated in a similar manner to those of the all-sky survey, but the
internal release to the
ASTRO-F team will commence during the survey period. Moreover, basic image
data (output from the
very early-stage processing) will also be released with the associated 
pipeline tools,
astronomical calibration
tools and tables, and scientific analysis tools which will be developed by
the data analysis team
for the NEP and the LMC surveys. These catalogues, the basic image data,
and the tools will also be
opened to the public approximately 1 year after the internal release.

\section{Observation Plan}

  \subsection{Mission phases}

Observations with ASTRO-F can be divided into the following phases:
\begin{itemize}
\item{PV phase [ L(launch)+1 -- L+2months ] : observations for performance 
verification
of the whole system.}
\item{Phase 1 [ L+2 -- L+8months ] : primarily for the all-sky survey. 
However, pointing
observations toward fields close to the ecliptic poles are planned on 
orbits passing
  through the SAA.  }
\item{Phase 2 [ L+8 -- L+18months(boil-off of liquid He) ] : supplemental 
all-sky survey for the area not sufficiently covered in phase-1, and pointing 
observations at any position on the sky.}
\item{Phase 3 [after boil-off of liquid He ] : pointing observations with 
only the near infrared channel of the IRC, which will work after the liquid He has boiled off.
 The duration of this phase is uncertain but is limited by the life of the mechanical coolers. }
\end{itemize}

Not only the large-area surveys described in the previous section, but also 
the following observation programs are planned:

\begin{itemize}
\item{Mission Program (MP): an organized program of pointing observations 
planned by the ASTRO-F science working groups (ASTRO-F team plus collaborators 
from around the world).
The targets are spread over large areas and cannot be included in the NEP and 
LMC surveys.}
\item{Open-time Program (OP): a program of pointing observations whose 
observing time will be open to Japan, Korea, and ESA communities. 30\% of 
pointing observation opportunities in phases 2 and 3 will be allocated to OP. }
\item{Director's Time: observing time allocated to the Project Manager, 
including the ``Target of Opportunities" time.
Less than 5\% of the total observation has been considered so far.}
\item{Calibration Time: for the long-term maintenance and performance 
verification for each instrument, calibration observations should be made 
at regular periods.}
\end{itemize}

  \subsection{Observation scheduling : simulation with trial models}

Since ASTRO-F is not an observatory but a sky surveyor for specific purposes, 
detailed observation plans especially for phases before the boil-off of the 
liquid helium have to be determined before the launch. This means we have to 
perform detailed pre-launch simulations of the time allocation and the 
spacecraft operations during the mission. 
The purpose of the simulation is to check the all-sky survey accomplishment
and to obtain realistic numbers for the pointing opportunities.

\begin{figure}[htbp]
\begin{center}
\includegraphics[width=14cm,clip]{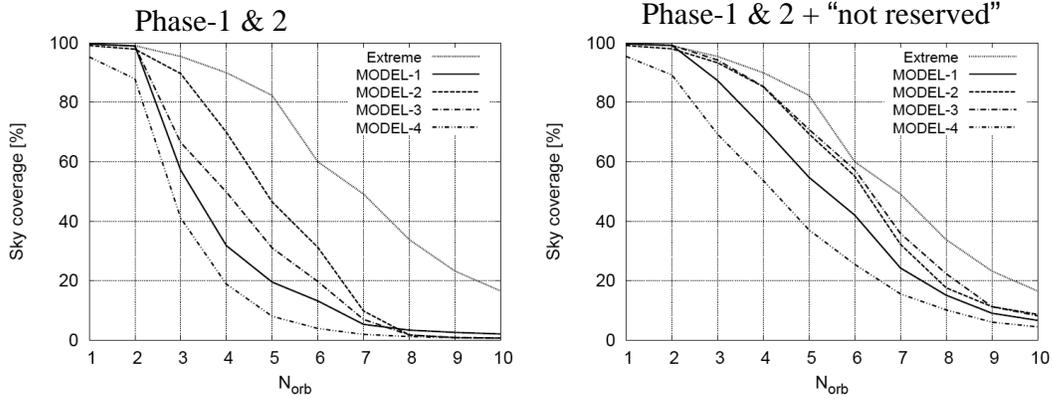}
\caption{All-sky survey accomplishment ratio for the trial models
described in Table~\ref{tab_models}. (Left) for all observation time allocated
for the survey in Phase-1 \& 2. (Right) in case of also including the 
``not-reserved" duration of time.}
\label{norbarea}
\end{center}
\end{figure}

As a first step, we have made a simple simulation of the time allocation
during phases 1 and 2 for four models with different allocation policies 
described in Table~\ref{tab_models}.
Here we define a ``SAA orbit" as an orbit that starts immediately after the 
spacecraft passes through the SAA. 
A provisional target list has been provided by summing requests from the
large area surveys and MP proposals. The total number of requested pointing
observations is 11,349. For simplicity we
ignore the detailed observational operation of the focal-plane instruments,
and do not consider any scientific priority in the target list. These models 
are characterized by the Phase 1 policy: Model 1 has no pointing observations
while Model 2 has pointing observations for regions at any ecliptic latitude
during an SAA orbit. In Models 3 and 4 the NEP and LMC pointing surveys
are made using the SAA orbits, but at lower ecliptic latitude no pointing 
observations are made, the orbit segments between the NEP and the LMC 
being used for the all-sky survey. The capability to maneuver 
the telescope in the cross-scan direction up to $\pm$1deg is
very useful to overcome conflicts in the target list. The resulting 
number of pointing opportunities allocated is shown in the last column of 
Table~~\ref{tab_models}. The number of pointings
allocated vary significantly between 5 and 10 thousand amongst the models.
The resulting all-sky survey coverage is shown in Figure~\ref{norbarea}. 
Here, we define as an index of survey
accomplishment ``$N_{\rm orb}$", the number of the scan paths in all-sky survey 
mode at a given point on the sky\footnote{$N_{\rm orb}$ is slightly different
from the definition of {\it HCON} of IRAS, because scan paths counted as 
$N_{\rm orb}$ are not always successive ones.}.  The ``not-reserved" duration means that there
is no appropriate target (for pointing or even for a survey scan with $N_{\rm orb} 
< 2$). However, by default, such times should be used for the survey, 
then the resulting sky coverage is shown in Figure~\ref{norbarea} (right).  
 The all-sky survey covers 
$>$90\% of the sky with $N_{\rm orb} \ge 2$ in all
of the models. Also more than 10\% of the sky is covered with $N_{\rm orb} 
\ge 6$: the medium deep survey in Figure~\ref{deptharea} (right).
To generate a reliable point source catalog and images from the 
all-sky survey, allocations to the survey observations for the area of
$N_{\rm orb} < 2$ should be prior to the pointing observations. At the
moment this simulation is still preliminary, since no such priority is
considered. 
Further investigation of the scheduling is necessary and is on-going, 
also considering the dead pixels of the FIS and cosmic-ray hits,
since these effects degrade the all-sky survey completeness.

\section{Summary}

We have described the updated pre-flight
sensitivities as well as as the observation plan  with
ASTRO-F, the first Japanese satellite mission dedicated for large
area surveys in the infrared. The primary
objective of the ASTRO-F mission is to make an all-sky survey in 6 wavebands
covering 9-160$\mu$m. Large-area {\it Legacy} surveys in pointing mode
will be also performed over large areas in the NEP and LMC regions. Based on a preliminary 
simulation of
the time allocation for both the all-sky survey and the pointing observations, we 
discuss the accomplishment
of the survey, and the realistic number of pointing opportunities.

The ASTRO-F project is managed by the Institute of Space and Astronautical 
Science (ISAS) of JAXA,
in collaboration with universities and institutes in Japan, Korea, UK, and 
the Netherlands.
The authors are deeply grateful to all the members of the ASTRO-F project 
for their help
and support. This work is partly supported by Grant-in-Aids for Scientific 
Research from the
JSPS.

% Bibliographic references with the natbib package:
% Parenthetical: \citep{Bai92} produces (Bailyn 1992).
% Textual: \citet{Bai95} produces Bailyn et al. (1995).
% An affix and part of a reference:
%   \citep[e.g.][Ch. 2]{Bar76}
%   produces (e.g. Barnes et al. 1976, Ch. 2).

\vskip 12pt

%%%%%%%%%%%%%%%%%%%%%%%%%%%%%%%%%%
%%%%% Figures and Tables   %%%%%%%
%%%%%%%%%%%%%%%%%%%%%%%%%%%%%%%%%%
%%%%% tab_imaging %%%%%%%%%
\begin{table}[h]
\caption{ASTRO-F waveband characteristics for broad-band imaging, used both
in survey and pointing. $\lambda_{\rm c}$ (in unit of $\mu$m) is
band center wavelength, and $\Delta \lambda$ is effective band-width, both of which 
are defined for $\nu F_{\nu} = \lambda F_{\lambda} = const.$ spectral sources.}
\label{tab_imaging}
\begin{center}
\begin{tabular}{cccccccccc} \hline
  band              &  N2  &  N3  &  N4  &  S7  &  S9W  &  S11  &  L15  & L18W  &  L24  \\ \hline
  $\lambda_{\rm c}$ & 2.4  &  3.2 &  4.1 & 7.3  &  9.1  & 10.7  & 15.7  & 18.3  &  23.0 \\
  $\Delta \lambda$  & 0.68 &  1.1 &  1.2 & 2.6  &  4.3 &   4.7  &  6.2  &  10.  &   5.4 \\ \hline
\end{tabular} \vskip 3pt
\begin{tabular}{ccccc} \hline
  band              &  N60  & WIDE-S & WIDE-L & N160 \\ \hline
  $\lambda_{\rm c}$ &   65  &    90  &   140  &  160  \\
  $\Delta \lambda$  &   22  &    38  &    52  &  34  \\ \hline
\end{tabular}
\end{center}
\end{table}

%%%% tab_survey %%%%%%%%
\begin{table}[h]
\caption{Flux limit(one-scan, 5$\sigma$) and size of the spatial pixel
for the all-sky survey.}
\label{tab_survey}
\begin{center}
\begin{tabular}{ccccccc} \hline
band                & S9W &  L18W & N60  & WIDE-S & WIDE-L & N160   \\ \hline
Point Source [mJy]  &  80 &  130  & 600  &    200 &    400 & 800    \\
Diffuse [MJy/sr]    &  25 &  32   &  7.5 &    2.5 &     2  &  3     \\
pixel size[arcsec]  & 9.6 &  9.6  &  30  &    30  &    50  &  50    \\ \hline
\end{tabular}
\end{center}
\end{table}

%%%% tab_pointing %%%%%%
\begin{table}[h]
\caption{Flux limit(5$\sigma$) at ecliptic poles for one pointing observation.}
\label{tab_pointing}
\begin{center}
\begin{tabular}{cccccccccc} \hline
band                    & N2    &  N3   &  N4   &  S7   &  S9W  &  S11  &  L15  & L18W  & L24   \\ \hline
Point Source [$\mu$Jy]  & 7.9   &  3.7  &  7.2  &  33   &   26  &   37  &   68  &   87  & 180   \\
Diffuse [MJy/sr]        & 0.032 & 0.013 & 0.022 & 0.059 & 0.045 & 0.062 & 0.093 & 0.094 & 0.16  \\ \hline
\end{tabular}
\end{center}
\end{table}

%%%% tab_slow %%%%%%
\begin{table}[h]
\caption{Flux limit(5$\sigma$) for one pointing observation with
slow-scan(15"sec$^{-1}$).}
\label{tab_slow}
\begin{center}
\begin{tabular}{ccccccccccc} \hline
band                &  S7  &  S9W  &  S11  &  L15  & L18W & L24  &  N60  & WIDE-S & WIDE-L & N160  \\ \hline
Point Source [mJy]  &  11  &   7   &  13   &   16  &  20  & 53   &   69  &   15   &   12   &  24   \\
Diffuse [MJy/sr]    &   4  &   2   &   4   &    4  &   4  &  9   &   1.4 &  0.36  &  0.03 & 0.05 \\ \hline
\end{tabular}
\end{center}
\end{table}

%%%% tab_models %%%%%%
\begin{table}[h]
\caption{Trial models for time-allocation simulation. The last column shows 
the resulting number
of pointing opportunities (phase-1 + phase-2).}
\label{tab_models}
\begin{center}
\begin{tabular}{lllc} \hline
Model    &  Phase-1                 &  Phase-2                      &  No. 
of pointing assigned  \\ \hline
Model-1  &  Survey on all orbits,   & Survey of $N_{\rm orb} < 2$   &  7857 
\\
          &  no pointing.            & is prior to pointing.         &  (0 
+ 7857)       \\ \hline
Model-2  &  pointing on SAA orbits, & Survey of $N_{\rm orb} < 2$   &  6182 
\\
          &  survey on others.       & is prior to pointing.         & 
(2965 + 3217)    \\ \hline
Model-3  &  pointing in NEP, LMC on & Survey of $N_{\rm orb} < 2$   &  5451 
\\
          &  SAA orbits, survey over & is prior to pointing.         & 
(1558 + 3893)    \\
          &  rest of time.           &                               & 
\\ \hline
Model-4  &  pointing in NEP, LMC on & Pointing is prior to survey,  &  9795 
\\
          &  SAA orbits, survey over & then survey if assignable.    & 
(1558 + 8237)    \\
          &  rest of time.           &                               & 
\\ \hline
\end{tabular}
\end{center}
\end{table}

%%%%% Figure Orbit %%%%%

%%%% Figure Spacecraft & Fov %%%%

%%%% Figure NEP and SEP (order to Imai) %%%%

%%%%%% Figure Flux Limit(survey) %%%%

%%%%%% Figure Depth vs Area %%%%

%%%%%% Figure Norb vs Area %%%%

\end{document}